\documentstyle[sprocl,epsf]{article}

\def\NPB{{\em Nucl. Phys.} B}
\def\PLB{{\em Phys. Lett.}  B}

\def\PRD{{\em Phys. Rev.} D}

\def\be{\begin{equation}}
\def\ee{\end{equation}}
\def\bea{\begin{eqnarray}}
\def\eea{\end{eqnarray}}
\def\Vec#1{\mbox{\boldmath $#1$}}

\def\itmb{\begin{itemize}}
\def\itme{\end{itemize}}
\def\enmb{\begin{enumerate}}
\def\enme{\end{enumerate}}
\def\eqnb{\begin{equation}}
\def\eqne{\end{equation}}
\def\eqab{\begin{eqnarray}}
\def\eqae{\end{eqnarray}}
\def\dis{\displaystyle}

\begin{document}

\title{NUMERICAL TESTS OF THE KUGO-OJIMA\\
 COLOR CONFINEMENT CRITERION}

\author{Hideo NAKAJIMA}

\address{Department of Information science, Utsunomiya University, 
Utsunomiya,\\ 321-8585 Japan\\E-mail: nakajima@is.utsunomiya-u.ac.jp}

\author{Sadataka FURUI }

\address{School of Science and Engineering, Teikyo University, 
Utsunomiya,\\ 320-8551, Japan\\E-mail: furui@dream.ics.teikyo-u.ac.jp}


\maketitle\abstracts{The Kugo-Ojima color confinement criterion,
which is based on the BRST symmetry of QCD is
numerically tested by the lattice Landau gauge simulation.
We first discuss the BRST symmetry on the lattice and the Gribov
copy problem. The horizon function defined by Zwanziger based
on the $U$-linear definition of the gauge field and that based  
on $\log U$ are compared.
The parameter $u$ which is expected to be $-1\delta^a_b$ in the
continuum theory was found to be $-0.7\delta^a_b$ in the strong
coupling region.  There are about 20\% difference between the two
horizon functions, but in the weak coupling region, the expectation 
value of the horizon function is suggested to be zero or negative.}

\section{Introduction}
There are essentially two aspects in the manifestation of color confinement in
QCD. One aspect is the linear potential between quarks, 
which was conjectured by Gribov as a consequence of an
enhancement of the singularity of the ghost propagator\cite{Gv} due to the 
restriction on the gauge field $A$.  

Another aspect is the absence of free single colored particle state in the 
asymptotic Hilbert space, which culminates in the Kugo and Ojima color 
confinement criterion based  on the BRST(Becchi-Rouet-Stora-Tyutin) symmetry:
i.e. in the Landau gauge, a coefficient in the two-point function produced by 
the ghost, the antighost and the gauge field becomes $-\delta_a^b$, where 
$a$ and $b$ specify the color in the adjoint representation. Analytical 
calculation of this value is extremely difficult and so far no verification 
was performed.

In 1994,  Zwanziger's developed a lattice QCD theory for Gribov ambiguity.
He claimed that, if the restriction to the fundamental modular region is
 achieved, the gluon propagator at 0 momentum should vanish in the 
continuum limit. 
In the Lattice QCD test, we address the following 
problems:  whether the gluon propagator is infrared finite, how singular the ghost propagator is in the infrared region, and whether the Kugo-Ojima color confinement criterion\cite{KO} is satisfied.

\section{The lattice simulation of the Landau gauge QCD and the Gribov problem}
\subsection{The path integral formulation of the gauge fixed theory in the
presence of the Gribov copy (Fujikawa, Hirschfeld)}

 First we give a brief review of the path integral formulation of the
gauge fixed theory by Fujikawa\cite{Fuj} and Hirschfeld\cite{Hir}. They discussed a possible
situation which may give a way out of the gauge fixing degeneracy problem
(the Gribov problem).
Let us define a gauge unfixed partition function as,
\begin{equation}
Z=\int dU e^{-\beta S(U)}.
\end{equation}
In derivation of the Faddeev-Popov(Faddeev-Popov) formula of the gauge $f(U)=0$,
one considers the determinant function $\Delta (U^g)$,
\[
\Delta (U^g)=\det \left(\displaystyle {\partial f(U^g)\over \partial U^g}
\displaystyle {\partial U^g\over \partial g}\right )
=\det \left(\displaystyle {\partial f(U^g)\over \partial A^g}
\displaystyle {\partial A^g\over \partial g}\right )
\] 
where $A_\mu=A_\mu(U)$, and one finds that $\Delta(U^g)$ is indeed
a function of $U^g$, and looks at the integral on the gauge orbit,
$U^g$,
\[
N(U)\equiv\int dg \Delta (U^g) \delta(f(U^g)).
\]
Obviously $N(U)$ is an orbit function, that is, $N(U^g)=N(U)$.
At the intersection points of the gauge orbit $U^g$ with the surface $f(U)=0$,
i.e., $g=g_i(U)$, Gribov's copies, the above delta function is transcribed
to give
\[
N(U)=\int dg \sum_i
\displaystyle{\Delta (U^g) \over |\Delta (U^g)|}\delta(g-g_i(U))
=\int dg \sum_i sign(\Delta (U^g))\delta(g-g_i(U)).
\label{EXX1}
\]
These delta functions contribute $0$ or $\pm 1$, and thus, if the orbit
function $N(U)$ is non vanishing over all orbits, $N(U)\ne 0$, then in
use of the identity
\[
1=\displaystyle{1\over N(U)}\int dg \Delta (U^g) \delta(f(U^g)),
\]
the standard FP procedure applies, and one factors out the
gauge volume and
obtains the formula
for expectation values of functions, $F(U)$,
\begin{equation}
\langle F(U)\rangle|_{f\ gauge}=
\displaystyle{\int dU \Delta (U)\delta (f(U))F(U)e^{-\beta S(U)}/N(U)
\over
\int dU \Delta (U)\delta (f(U))e^{-\beta S(U)}/N(U)}.
\label {EX1}
\end{equation}
This formula allows the BRST formulation as in the following, and also
derives a natural simulation algorithm of the gauge fixed theory.
\subsection{The simulation algorithm for the gauge fixed
 theory (Mandula-Ogilvie) and the Gribov problem}
Multiplying by the gauge volumes, the denominator and the numerator of the
formula (\ref{EX1}), respectively, one recovers the $Z$ in the denominator,
and has in the numerator,
\begin{equation}
\int dU \int dg \Delta (U^g)\delta (f(U^g))F(U^g)e^{-\beta S(U)}
/N(U).
\end{equation}
In use of (\ref{EXX1}), one obtains the numerator as
\begin{equation}
\int dU \sum_i sign(\Delta(U^{g_i}))F(U^{g_i})e^{-\beta S(U)}/N(U)
\end{equation}
where $U^{g_i}$ is the i-th Gribov copy on the orbit $U^g$. This gives us
the algorithm in the simulation\cite{MO}
\begin{equation}
\langle F(U)\rangle|_{f\ gauge}
=\displaystyle{1\over Z}\int dU \sum_i \displaystyle{sign(\Delta(U^{g_i}))F(U^{g_i})\over N(U)}e^{-\beta S(U)}
=\langle \bar F(U)\rangle
\end{equation}
where the last averaging $\langle\rangle$ is that of simulation,
i.e., that with respect to the Boltzmann weight $e^{-\beta S}$, and
$\bar F(U)$ is a sign weighted average of $F(U^{g_i})$ on the
gauge orbit $U^g$. 

However, in case of gauge non invariant function $F(U)$, 
this formula would rather be interpreted as representing Gribov noise.
Thus a modification of the gauge such that one chooses a unique copy
among others on the orbit is favored, that is, {\bf a new gauge without 
Gribov copy}, and in that case, the above formula is
useful in practice.

\subsection{The BRST formulation and the Gribov problem}
The obvious standard FP formula allows BRST formulation,
\begin{equation}
\langle F(U)\rangle|_{f\ gauge}=\displaystyle{1\over Z'}
\int d\mu \exp( \delta \int \bar c f(U))F(U)e^{-\beta S(U)}/N(U),
\label {EX2}
\end{equation}
\begin{equation}
Z'=\int d\mu \exp( \delta \int \bar c f(U))e^{-\beta S(U)}/N(U),
\label {EX3}
\end{equation}
where $\delta$ stands for  BRST transformation,
and the measure is defined as
$d\mu=dUdBdcd\bar c,
\label {EX5}$
where $dB$ integration is performed on ${\Vec R}$ and $dcd\bar c$ is
suitablly defined differentiations with respect to Grassmann numbers, $c$ and
$\bar c$.

It was shown by Neuberger that if the gauge fixing function $f(U)$ is a smooth
function of compact variables $U$, then the expectation value
of gauge invariant function $F(U)$ becomes an indefinite form\cite{Neu},
\begin{equation}
\langle F(U)\rangle|_{f\ gauge}=\displaystyle{0\over 0},
\end{equation}
which implies that all Gribov copies contribute to give total cancellation,
in other words, the assumption that the above $N(U)$ is non vanishing on all
orbits, does not hold, and the formula (\ref{EX1}) is totally meaningless.\par
An essential point of his argument is as follows.
Let us consider a general expression
\begin{equation}
Z'(t)=\int d\mu \exp(t \delta \int \bar c f(U))G(U)
\label {EX8}
\end{equation}
with a gauge invariant function $G(U)$ as the Boltzmann weight. Then
one finds that
\begin{equation}
\displaystyle{dZ'\over dt}(t)=\int d\mu (\delta \int \bar c f(U))
\exp(t \delta \int \bar c f(U))G(U)
\label {EX9}
\end{equation}
can be written from nilpotency of $\delta$, i.e., $\delta^2=0$, and
from $\delta G(U)=0$ as,
\begin{equation}
\displaystyle{dZ'\over dt}(t)=\int d\mu \delta \left((\int \bar c f(U))
\exp(t \delta \int \bar c f(U))G(U)\right).
\label {EX10}
\end{equation}
If $W(U,B,c,\bar c)$ is an analytic function of the compact variables $U$, 
then one can show that
\begin{equation}
\int d\mu \delta W(U,B,c,\bar c)=0.
\label {EX11}
\end{equation}
Thus it follows that
$\displaystyle{dZ'\over dt}(t)=0.$
Together with $Z'(0)=\int \cdots d\bar c G(U)=0$, one finds that
$Z'(t)=0$, so $Z'(1)=0$. This concludes total cancellation of
the Gribov copies of gauges given by analytic gauge functions of $U$.
Thus one is forced to consider non analytic gauge functions of $U$ as
desired gauge functions. As a one dimensional $U(1)$ toy example
which avoids (\ref{EX11}),
one can consider  $U=e^{iA}$, $A={\rm Im} \log U$, where $A$ is not continuous
at $U=-1$, and with definition $\delta A=c$, one finds that
\begin{equation}
\int dUdBdcd\bar c\ \delta( e^{-B^2}\bar c A)=
\int_{-\pi}^{\pi} dA\int _{-\infty}^{\infty}dBdcd\bar c\ e^{-B^2}c\bar c\ne 0.
\label {EX12}
\end{equation}
Although this problem is still open in the lattice BRST formulation,
use of non compact variables in gauge fixing functions may be helpful.

\subsection{The Landau gauge and the Gribov problem}
Now we focus on the Landau gauge in $SU(3)$ QCD, that is,
$f(U)=\partial_\mu A_{\mu}(U),$
where there are some options of definition $A_{\mu}(U)$ as 
\begin{enumerate}
\item $U$-linear one\cite{Zw};
$A_{x,\mu}=\displaystyle{1\over 2}(U_{x,\mu}-U_{x,\mu}^{\dag})|_{traceless\ part},$

\item
use of exponential map\cite{NF}.
$U_{x,\mu}=\exp{A_{x,\mu}},\ \ \ A_{x,\mu}^{\dag}=-A_{x,\mu},$
where absolute values of all eigenvalues of $A_{x,\mu}$ do not exceed $4\pi/3$.
\end{enumerate}
In the latter definition, $A_{x,\mu}(U)$ is not analytic with
respect to compact variable $U_{x,\mu}$ contrary to the former one.
In both cases the Landau gauge, $\partial A^g=0$, can be characterized in use
of optimizing functions $F_U(g)$ of $g$, such that $\delta F_U(g)=0$ for
any $\delta g$.
\begin{enumerate}
\item $U$-linear definition;
$F_U(g)={2N\over dV(N^2-1)}\sum_{x,\mu}\left(1-{1\over N}{\rm Re}\ {\rm tr} U^g_{x,\mu}\right).$
\item use of exponential map;
$F_U(g)=||A^g||^2={1\over dV(N^2-1)}\sum_{x,\mu}{\rm tr}
 \left({{A^g}_{x,\mu}}^{\dag}A^g_{x,\mu}\right)$,
where $N$ is the number of colors and $d$ is the dimension. 
It is noteworthy that 
$||A^g||^2$ is a continuous function of compact
variables $U^g$ in spite of non analytic property of $A_{x,\mu}(U^g)$.
\end{enumerate}

Gribov copy is generic phenomena in both definitions, there exist
a lot of local minima of $F_U(g)$ along the gauge orbit $U^g$.
Thus the naive Landau gauge loses its rigid basis
both in the theoretical and in the simulation view points for
examination of gauge non invariant quantities such as
gluon propagator, ghost propagator, etc. 

The arguments in the preceding subsections formally applies for a corresponding 
gauge function containing non analytic Heaviside function. The gauge-fixing algorithm in the simulation is required
to attain the {\bf absolute minimum} of the $F_U(g)$ along the gauge orbit.
But the global minimization is difficult in general and 
developing the efficient algorithm is still an open problem\cite{YN}.


\section{The Kugo-Ojima confinement criterion and the Gribov-Zwanziger's theory}
\subsection{Kugo-Ojima's theory}

A sufficient condition of the color confinement given 
by Kugo and Ojima\cite{KO} is that
 $u^{ab}$ defined by the two-point
function of  the FP  ghost fields, $c(x),\bar c(y)$,
and $A_\nu(y)$,
\begin{equation}
\int e^{ip(x-y)}\langle 0|T D_\mu c^a(x)g(A_\nu\times \bar c)^b(y)|0\rangle d^4x=
(g_{\mu\nu}-{p_\mu p_\nu\over p^2})u^{ab}(p^2)
\label{eq}
\end{equation}
satisfies $u^{ab}(0)=-\delta^{ab}$.

The corresponding Euclidian expression
can be calculated by
\begin{equation}
\displaystyle{1\over V}
\sum_{x,y} e^{-ip(x-y)}\langle  {\rm tr}\left({\lambda^a}^{\dag}
D_\mu \displaystyle{1\over \partial D}[-A_\nu\lambda^b] \right)_{xy}\rangle=
(\delta_{\mu\nu}-{p_\mu p_\nu\over p^2})u^{ab}(p^2),
\label{eqq}
\end{equation}
where $\lambda^a$ is a normalized antihermitian basis of Lie algebra, $V$ a 
lattice volume,
and
the ghost propagator is given by
\begin{equation}
\langle c^a_x\bar c^b_y\rangle
=\langle{\rm tr}\left({\lambda^a}^{\dag}
\displaystyle{1\over \partial D}\lambda^b \right)_{xy}\rangle.
\label{eqq1}
\end{equation}

\subsection{Zwanziger's theory}
 The {\bf fundamental modular region} $\Lambda$ is specified by the
absolute minimum along the gauge orbits in the {\bf Gribov region} $\Omega$.
\begin{equation}
\Lambda=\{A|\|A\|^2={\rm Min}_g\|A^g\|^2\}, \qquad
\Lambda\subset \Omega=\{A|-\partial { D}\ge 0\ ,\ \partial A=0\}\ \ .
\label{NORM9}
\end{equation}

 Zwanziger relaxes the periodicity restriction on the gauge transformation
$g$, and imposes larger periods than the original. 
Then some two points in the fundamental modular region $\Lambda$
may be bridged to be Gribov copies of each other, and one of them is not
the absolute minimum of the minimizing function along the gauge orbit anymore.
Surviving points as the absolute minimum consist of {\bf core region} 
$\Xi$ ( $\Xi\subset\Lambda$).
In the so defined core region $\Xi$, a horizon function $H(U)$ given below is negative.
 
 The Horizon function is defined as follows.
Let two point tensor ${G_{\mu\nu xy}}^{ab}$ be
\begin{equation}
{G_{\mu\nu xy}}^{ab}
=
 {\rm tr}\left({\lambda^a}^{\dag}
D_\mu \displaystyle{1\over -\partial D}(-D_\nu)\lambda^b\right)_{xy}.
\label{eqq3}
\end{equation}
Then $H(U)$ is given as
\begin{equation}
H(U)=\sum_{x,y,a}{G_{\mu\mu xy}}^{aa}-(N^2-1)E(U)
\end{equation}
where $E(U)$ reads as follows;
\enmb
\item in $U$-linear version,
$E(U)=\sum_l{1\over N}{\rm Re}\ {\rm tr} U_l$,
\item in $A=\log U$ version,
$E(U)=\dis{{1\over N^2-1}\sum_{l,a}}{\rm tr}\left ({\lambda^a}^\dag S({\cal A}_l)\lambda^a\right)$,

where ${\cal A}_l=adj_{A_l}$, and 
$S(x)=\dis{{x/2\over {\rm th}(x/2)}}$.
\enme
Let us define an average tensor $G_{\mu\nu xy}$ be
$G_{\mu\nu xy}\delta^{ab}=\langle G_{\mu\nu xy}^{ab}\rangle,$
provided color symmetry is not broken.
One sees that a Fourier transform of the average tensor,
takes a form
\begin{equation}
G_{\mu\nu}(p)\delta^{ab}=
\left(\dis{e\over d}\right)\displaystyle{p_\mu p_\nu\over p^2}\delta^{ab}
-\left(\delta_{\mu\nu}-\displaystyle{p_\mu p_\nu\over p^2}
\right)u^{ab},
\label{GMN}
\end{equation}
where
$e=\langle E(U)\rangle/V.$,
and that it is related with the horizon function as
\begin{equation}
\displaystyle{\langle H(U)\rangle\over V}
=(N^2-1)\left[\lim_{p\to 0}
G_{\mu\mu}(p)-e\right].
\label{INFZW}
\end{equation}

 He defined the {\bf augmented core region} $\Psi=\{U : H(U)\le 0
\}\cap \Omega$ ( $\Xi\subset\Psi\subset \Omega$ ). $\Psi$ and $\Lambda$ are 
qualitatively similar, and he defined the partition function $Z_{\Psi}$
in the path integral in use of the corresponding Landau gauge function 
$f_{\Psi}(U)$, and concluded\cite{Zw} in the infinite volume limit that
$\lim_{V\to \infty}\displaystyle{\langle H(U)\rangle/ V}=0$. 
 Putting Kugo-Ojima parameter as
$u^{ab}(0)=-\delta^{ab}c,$
one finds from (\ref{GMN}), (\ref{INFZW}), that
\[
\left(\dis{e\over d}\right)+(d-1)c-e=(d-1)\left(c-\dis{e\over d}\right)=0,
\]
which is called horizon condition.
Since we can measure $c$ and $e$ by the lattice simulation, we can check
to what extent Zwanziger's horizon condition holds in
our simulation. With respect to the value $e/d$, note that the
classical vacuum is characterized by $e/d=1$.

\section{The numerical results}

\subsection{The formulation of lattice Landau gauge fixing}

 We define the gauge field\cite{NF}
on links as an element of $SU(3)$ Lie algebra as,

\begin{equation}
e^{A_{x,\mu}}=U_{x,\mu}\ \ ,\ \ \ {\rm where}\ \ A_{x,\mu}^{\dag}=-A_{x,\mu}.
\label{DFGAUGE}
\end{equation}

 We perform the gauge transformation as
$e^{A^g_{x,\mu}}=g_x^\dagger  e^{A_{x,\mu}} g_{x+\mu}$
and define
$\displaystyle |\partial A|=Max_{x,\mu,a} |\partial A^a_{x}| ,
\|\partial A \|^2={1\over V(N^2-1)}\sum tr \partial A_{x}^\dagger \partial A_{x}$.

 The Landau gauge is realized by minimizing $\|A^g\|^2$ via a gauge
transformation $g^\dagger U g$, where $g=e^\epsilon$.
In order to obtain $\epsilon$, we switch the following two methods, depending
on the current value of $|\partial A|$ in comparison to some critical 
parameter $|\partial A|_{cr}$.  

\begin{enumerate}
\item When $|\partial A|>|\partial A|_{cr}$,
$\epsilon_x={\eta'\over \|\partial A\|}\partial A_x$
with suitable parameter $1<\eta'<2.2$ 

\item
When $|\partial A|<|\partial A|_{cr}$,
 $\epsilon=(-\partial_\mu D_\mu(A))^{-1}\eta\partial A $
where $1<\eta<2$ is a parameter.
\end{enumerate}

 The restriction to the fundamental modular region is not always achieved. 
But, we observed that the obtained norm $\|A\|$ is larger or smaller than that 
obtained after the smeared gauge fixing\cite{HdF} within 1\% accuracy.

\subsection{The Kugo-Ojima two-point function and the ghost propagator}

 The FP operator is
\begin{equation}
{\cal M}[U]=-(\partial\cdot D(A))=-(D(A)\cdot\partial),
\end{equation}
where the definition of the gauge field is adopted as $U=e^A$,
and the lattice covariant derivative $D_{\mu}(A)=\partial_{\mu}+Ad(A_{\mu})$
is given in ref.\cite{NF}.

  The inverse, ${\cal M}^{-1}[U]=(M_0-M_1[U])^{-1}$, is calculated
 perturbatively by using the Green function of the  Poisson equation 
$M_0^{-1}=(-\partial^2)^{-1}$  and $M_1=\partial_\mu Ad (A_\mu(x))$.

 The ghost propagator (\ref{eqq1}) is infrared divergent and its singularity can be
parameterized as $\displaystyle p^{-2(1+\alpha)}$, where $\displaystyle p^2=\sum_{k\mu} (4 \sin^2{ \pi k_\mu\over L})$, $(-L/2 <k_\mu\le L/2)$. 
It depends on
 $\beta$ slightly, but its finite-size effect is small\cite{SS}.
These qualitative features are in agreement with the 
analysis of the Dyson-Schwinger equation\cite{SHA}.

\begin{figure}[htb]
\begin{minipage}[b]{0.47\linewidth}
\begin{center}
\epsfysize=100pt\epsfbox{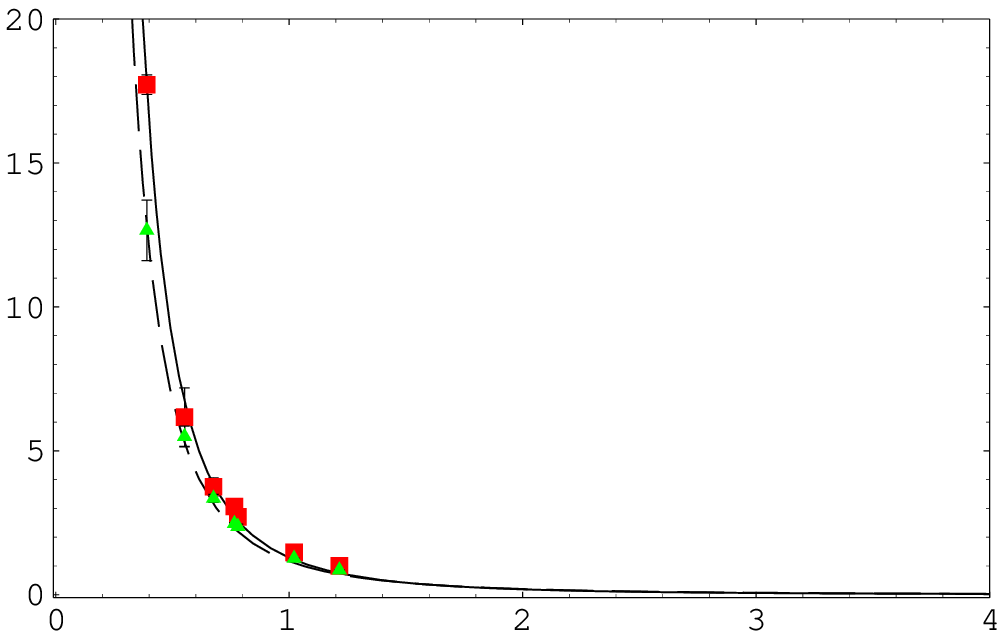}
\caption{The ghost propagator as function of the lattice momentum. 
The data are $\beta=5.5$(box) and $6.0$(triangle), $16^4$. The fitted curve is $1.287/p^{2.779}$ for $\beta=5.5$ and $1.162/ p^{2.545}$ for $\beta=6.0$ (dashed).}
\end{center}
\end{minipage}
\hfil
\begin{minipage}[b]{0.47\linewidth}
\begin{center}
\epsfysize=100pt\epsfbox{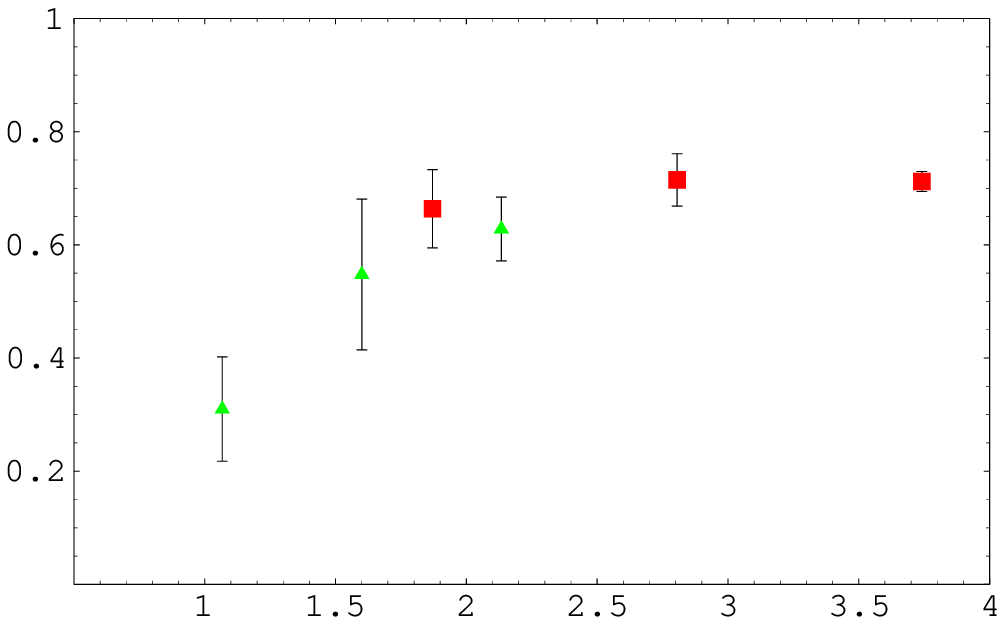}
\caption{The Kugo-Ojima parameter $|u^a_a|$ as the function of the spatial 
extent of the lattice $a L (fm)$.  The data are $\beta=6$ (triangle), and 
$\beta=5.5$ (box) $8^4, 12^4, 16^4$ from left to right, respectively.}
\end{center}
\end{minipage}
\end{figure}

We  measured the lattice version of  $|u^a_b(0)|$ on $8^3\times 16, 8^4$ 
$12^4$ and $16^4$ for $\beta=5.5$ and $6$. 
When $\beta=6$ and the lattice size is small, the Polyakov loop distribution 
deviates from the uniform distribution. In this case, we perform the 
${\Vec Z}_3$ rotation by multiplying the global phase $e^{\pm {2\pi\over 3}i}$
such that the distribution concentrate around one angle, before we measure
the Kugo-Ojima two-point function.

We obtained that $u^a_b(0)$ is consistent with $-c\delta^a_b,\ c=0.7$ in
$SU(3)$ quenched simulation, $\beta =5.5$, on $8^4, 12^4$ and $16^4$.

\subsection{The gluon propagator }

 The gluon propagator is infrared finite. We parameterized the zero-temperature
lattice data using the Stingl's Factorised Denominator Rational Approximant
(FDRA) method. The effective mass of the gluon in the analysis of $8^3\times 16$ is found to be about $600 MeV$.

The infrared finiteness is in accordance with the Kugo-Ojima color confinement
mechanism. 
 As stated in the their inverse Higgs mechanism theorem, 
if we have no mass less vector poles in all channels of the gauge field,
$A^a_{\mu}$, and if the color symmetry is not broken at all,
it follows that $1+u=0$.

\section{Discussion and conclusions}

We performed the first test of the Kugo-Ojima color confinement
criterion in lattice Landau gauge. We observed that the $8^4, 12^4$ and $16^4$
 lattice data. The data of $\beta=5.5$ indicates that $u=-0.7$, and those
of $\beta=6.0$ are smaller by about 10\%.

 In the Zwanziger's theory, the two-point function $G_{\mu\nu}(k^2)$
can be expressed in terms of the Kugo-Ojima two-point function as (\ref{GMN}).
Zwanziger's horizon condition\cite{Zw} in the infinite volume limit reads as
In terms of the Kugo-Ojima parameter $c$, the left hand side can be written as 
$(e/4)+3c$, and the horizon condition is that $c=e/d$. 
In the table below $e_1$
and $e_2$ stand for $e$ in our $16^4$ lattice simulation of the first
and the second option of the gauge fields, respectively.
If the gauge fixing is performed so that it brings the configuration into
the core region or the augmented core region and if the infinite volume
limit is considered somehow, then the legitimate check of the horizon
condition can be done. The core gauge fixing is, however, difficult, and
even impossible in general, which implies that the core gauge is literarily
 not the gauge, and thus we give the direct results in the table.

\begin{table}[htb]
\begin{tabular*}{\textwidth}{@{}l@{\extracolsep{\fill}}cccccccc}
\hline
    & $\beta$    & c  &   $G_{\mu\mu}(0)$ & $e_1/d$ &$e_2/d$& $1/\tilde Z_3$\\
\hline
 &5.5  & 0.712(18)  &  3.14(5)  & 0.783 &0.657 & 0.78\\
 &6.0  & 0.628(70)  &  2.88(17)  & 0.860(3) &0.693(1) & 0.86\\
\hline
\end{tabular*}
\end{table}

Although the renormalization factor of the ghost propagator $\tilde Z_3$ 
cannot be fitted  precisely, its inverse is numerically close to $e_1/d$.
Simulation data show in general that when $\beta$ becomes larger, $e$
becomes smaller, while $c$ has an opposite tendency. This fact itself
does not necessarily disprove the horizon condition, but
our preliminary data of $c$ which is calculated in the $A=\log U$ version
 already
gives the zero-intersection of $c-e_2/d$ in the increase of $\beta$
from $5.5$ to $6$.

This work was supported by KEK Supercomputer Project(Project No.99-46), and by JSPS, Grant-in-aid for Scientific Research(C) (No.11640251).

\end{document}